\documentclass[twoside,twocolumn,aps,secnumarabic,balancelastpage,amsmath,amssymb,hyperref=pdftex,superscriptaddress]{revtex4}
\usepackage[latin9]{inputenc}
\usepackage{xcolor}
\usepackage{mathtools}
\usepackage{bm}
\usepackage{amsmath}
\usepackage{graphicx}
\usepackage{braket}
\usepackage[unicode=true,
 bookmarks=false,
 breaklinks=false,pdfborder={0 0 1},backref=section,colorlinks=true]
 {hyperref}

\makeatletter
\@ifundefined{textcolor}{}
{%
 \definecolor{BLACK}{gray}{0}
 \definecolor{WHITE}{gray}{1}
 \definecolor{RED}{rgb}{1,0,0}
 \definecolor{GREEN}{rgb}{0,1,0}
 \definecolor{BLUE}{rgb}{0,0,1}
 \definecolor{CYAN}{cmyk}{1,0,0,0}
 \definecolor{MAGENTA}{cmyk}{0,1,0,0}
 \definecolor{YELLOW}{cmyk}{0,0,1,0}
}

\usepackage{graphics}

\@ifundefined{showcaptionsetup}{}{%
 \PassOptionsToPackage{caption=false}{subfig}}
\usepackage{subfig}
\makeatother

\begin{document}

\title{Fluctuations and Higgs mechanism in Under-Doped Cuprates : a Review}

\author{C. Pépin}

\affiliation{Institut de Physique Th\'eorique, Universit\'e Paris-Saclay, CEA, CNRS, F-91191 Gif-sur-Yvette, France.}

\author{D. Chakraborty}

\affiliation{Institut de Physique Th\'eorique, Universit\'e Paris-Saclay, CEA, CNRS, F-91191 Gif-sur-Yvette, France.}

\author{M. Grandadam}

\affiliation{Institut de Physique Th\'eorique, Universit\'e Paris-Saclay, CEA, CNRS, F-91191 Gif-sur-Yvette, France.}

\author{S. Sarkar}

\affiliation{Institut de Physique Th\'eorique, Universit\'e Paris-Saclay, CEA, CNRS, F-91191 Gif-sur-Yvette, France.}

\maketitle
\textbf{The physics of the pseudo-gap phase of high temperature cuprate
superconductors has been an enduring mystery in the past thirty years. The ubiquitous presence of the pseudo-gap phase in under-doped cuprates suggests that its understanding holds a key in unraveling the origin of high temperature superconductivity. In this paper, we review various theoretical approaches to this problem, with a special emphasis on the concept of emergent symmetries in the
under-doped region of those compounds. We differentiate the theories
by considering a few fundamental questions related to the rich phenomenology
of these materials. Lastly we discuss a recent idea of two kinds of
entangled preformed pairs which open a gap at the pseudo-gap onset temperature $T^{*}$
through a specific Higgs mechanism. We give a review of the experimental
consequences of this line of thoughts.}

\begin{figure}
\subfloat[]{\includegraphics[width=0.8\linewidth]{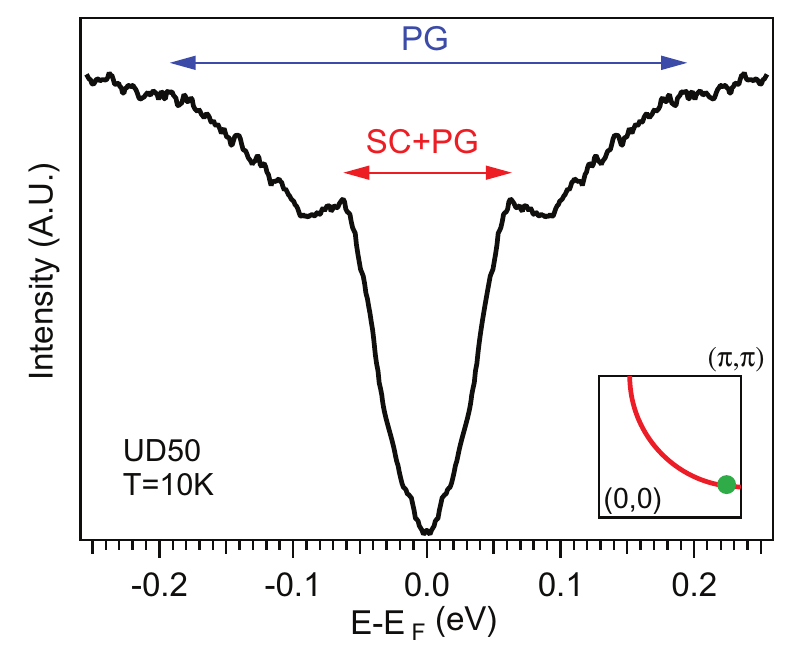}\label{Fig1a}

}

\subfloat[]{\includegraphics{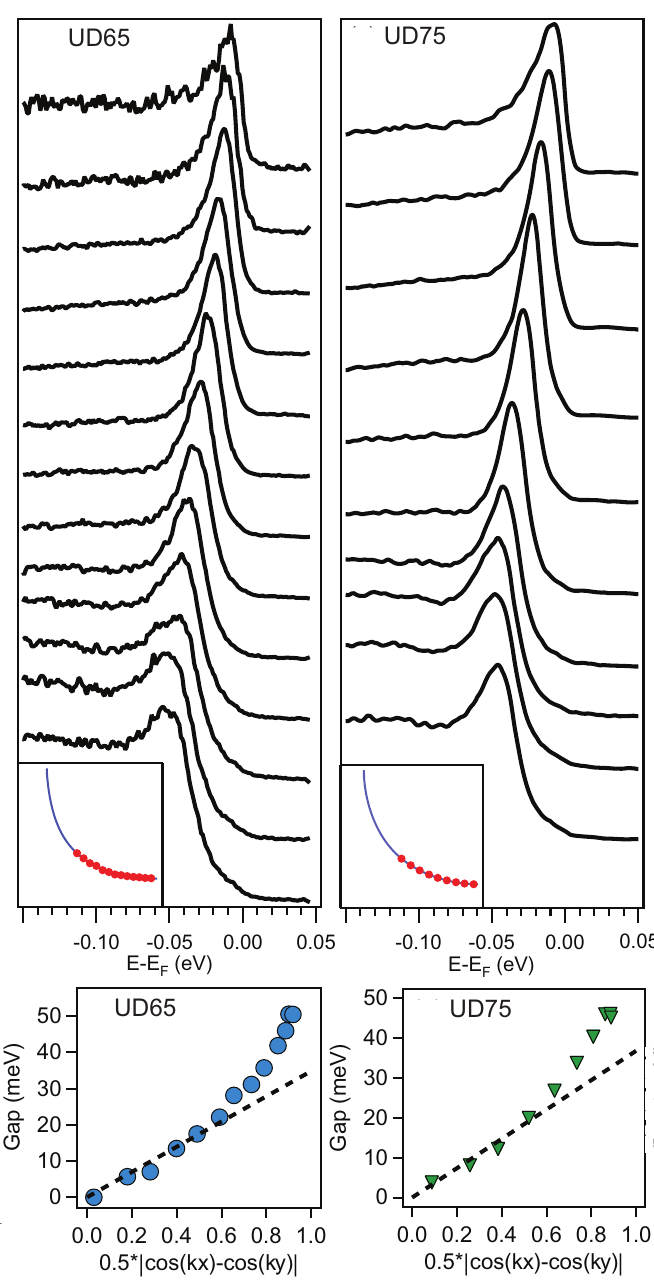}\label{Fig1b}

}

\caption{a) The anti-nodal gap measured by ARPES spectroscopy, which shows
two gaps, the first energy scale correspond to the PG scale $E^{*}$
whereas the second energy scale corresponds to the spectroscopic gap
$E_{spec}^{*}$. The data have been symmetrized with respect to the
energy $E_{F}$. b) Two sets of ARPES lines scanning the Fermi surface
of Bi1221 in the under-doped region. One sees that the quasi-particle
peak at $E=0$ present in the nodal region (the upper curves) evolves
into a Bogoliubov quasi-particle peak at finite $E$ in the anti-nodal
region (the lower curves). One can see how the Bogoliubov peak in the anti-nodal
region is still visible, whereas somewhat broadened. Data taken from
Ref. \cite{Vishik:2010fn}}
\end{figure}

The Pseudo-Gap (PG) state of the cuprates was discovered in 1989 \cite{Alloul89}, three years after the discovery of high temperature superconductivity in those compounds. It was first observed in nuclear magnetic resonance (NMR) experiments, in an intermediate doping regime $0.06<p<0.20$, as a loss of the density of states at the Fermi level \cite{Alloul89,Alloul91,Warren89} at temperatures above the superconducting transition temperature $T_c$. Subsequently, angle resolved photo emission spectroscopy (ARPES) established that a part of the Fermi surface is gapped in the Anti-Nodal Region (ANR) (regions close to $\left(0,\pm\pi\right)$ and $\left(\pm\pi,0\right)$ points) of the Brillouin zone, leading to the formation of Fermi `arcs'. Though the PG state shows behaviors of a metal, the appearance of Fermi `arcs' instead of a full Fermi surface results into the violation of the conventional Luttinger theorem of Fermi liquid theory. Furthermore, surface spectroscopies like ARPES (see e.g. \cite{Campuzano98,Campuzano99,Shen:2005ir,Vishik:2010fn,Vishik:2010tc}) and scanning tunneling spectroscopy \cite{Wise08,Hoffman02,Hamidian15a} show that the magnitude of the anti-nodal (AN) gap is unchanged when entering the superconducting (SC) phase below $T_{c}$ (see Fig. \ref{Fig1a}). The PG state persists up to a temperature $T^*$ which decreases linearly with doping. The AN gap is also visible in two-body spectroscopy, for example in the $B_{1g}$ channel of Raman scattering\cite{Benhabib:2015ds,Loret19,Loret:2017cv}, which shows that below $T_{c}$, the pair breaking gap of superconductivity follows the $T^{*}$ line with doping, in good agreement with the AN gap observed in ARPES. In contrast to the SC phase, the PG state is found to be independent of disorder or magnetic field. Despite of numerous invaluable experimental and theoretical investigations over the last three decades, various puzzles of the PG state remains to be solved.

In this paper, we review three different theoretical perspectives to the PG state of cuprate superconductors. We then focus on one specific theoretical approach where fluctuations protected by a specific Higgs mechanism are held responsible for the unusual properties of the pseudo-gap phase.

\section{Mott physics and strongly correlated electrons}

\begin{figure}
\includegraphics[width=0.9\linewidth]{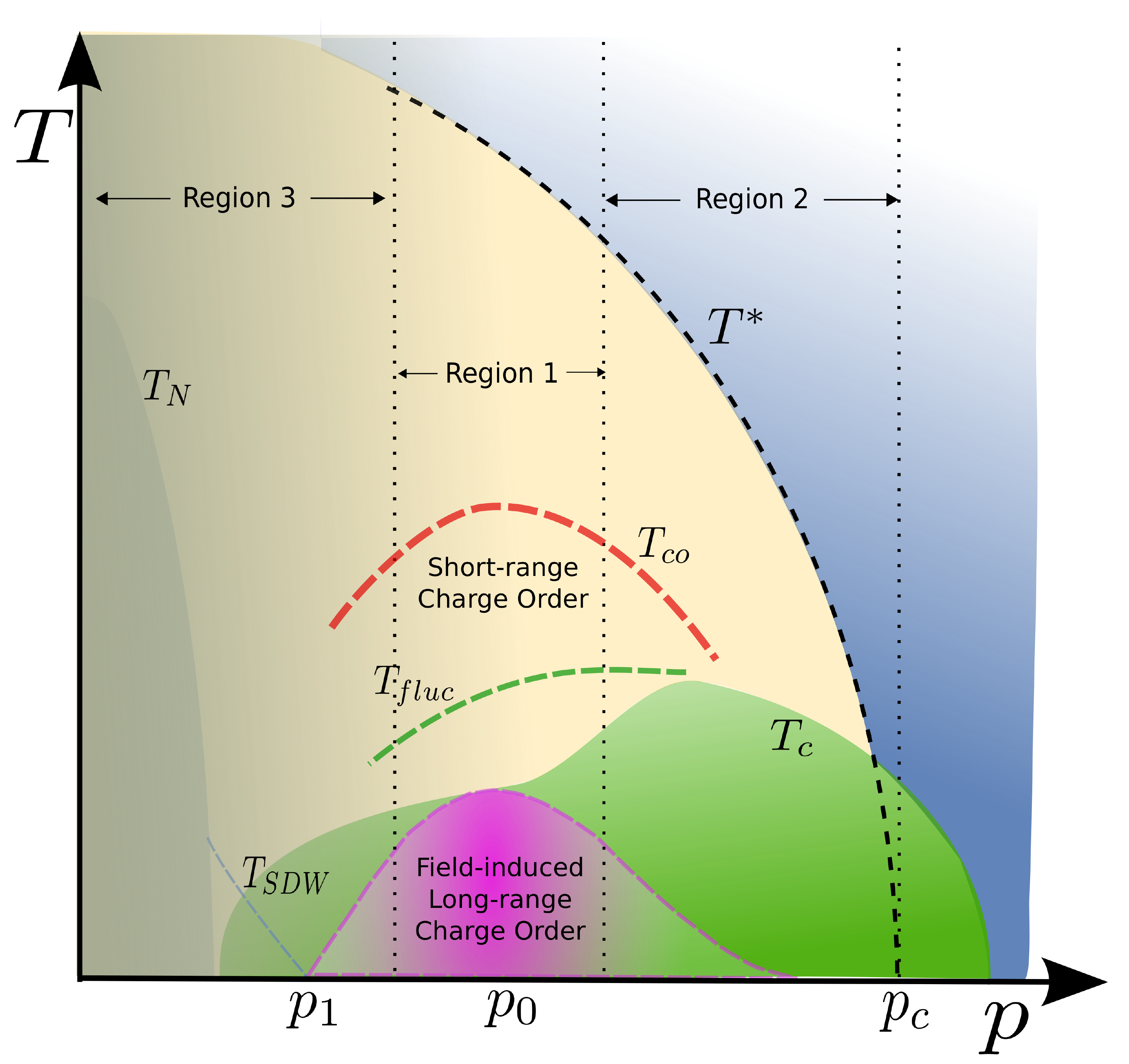}

\caption{\label{fig:The-phase-diagram}The schematic phase diagram of cuprates
superconductors as a function of hole doping \cite{Chakraborty18}. On
the left hand side, for $0<p<0.06$, the system has an AF phase. The
Green region below $T_{c}$ denotes the superconducting state. Below
$T_{co}$ the short range charge order is observed, whereas the purple
region is the charge order observed with application of a magnetic
field. $T^{*}$ is represented with a dashed line which encloses the PG phase marked with yellow. In Region 3 and Region 2 we have additional source of damping, respectively due to the approach of the Mott transition and the proximity to the PG quantum critical point.}
\end{figure}

An overall glance at the phase diagram of the cuprate superconductors
shows that superconductivity sets up close to an anti-ferromagnetic
(AF) phase transition, but also close to a Mott insulating phase (Fig.
\ref{fig:The-phase-diagram}). The presence of the metal-insulating
transition so close to superconductivity is unusual and led many theoreticians
to attribute the PG phase to a precursor of the Mott transition (a
few review papers \cite{Norman03,Lee06,Pines2002,Rice:2012eoa,SenthilLee09,Verret:2017he,Fradkin15,Alloul14,Scalapino:2006uw,LeHur:2009iw}).
The Coulomb interaction has a high energy scale $U=1eV$ which prohibits
the double occupancy on each site and induces strong correlations
between electrons. An effective Hamiltonian can be derived by integrating
out formally the Coulomb interaction, which leads to AF super-exchange
interaction associated with a constraint prohibiting double occupancy
on each site for conduction electrons (see e.g. \cite{Lee98,Lee92,Lee06})
and is given by 
\begin{equation}
H=\sum_{i,j}t_{ij}c_{i\sigma}^{\dagger}c_{j\sigma}+J\sum_{\left\langle i,j\right\rangle }S_{i}\cdot S_{j},\label{eq:1}
\end{equation}
where $c_{i,\sigma}^{\dagger}$ ($c_{i,\sigma}$) is a creation (annihilation)
operator for an electron at site $i$ with spin $\sigma$, $\bm{S}_{i}=c_{i,\alpha}^{\dagger}\bm{\sigma}_{\alpha,\beta}c_{i,\beta}$
is the spin operator at site $i$ ($\bm{\sigma}$ is the vector of
Pauli matrices), $t_{ij}=t_{ji}$ and $J$ is the spin-exchange interaction.
No double occupancy constraint is ensured by taking $n_{i}\le1$ with
$n_{i}=\sum_{\sigma}c_{i,\sigma}^{\dagger}c_{i,\sigma}$ being the
number operator. Theories then associate the formation of the PG phase
as a precursor of the Mott transition, involving some fractionalization
of the electron due to the constraint $n_{i}\le1$. The typical and
simplest realization of such a program is to consider that the electron
fractionalizes into ``spinons'' ($f_{i\sigma}^{\dagger}$) and ``holons''
($b_{i}$) subject to a local U(1) or gauge symmetry group, 
\begin{align}
c_{i\sigma}^{\dagger} & =f_{i\sigma}^{\dagger}b_{i},\label{eq:2}
\end{align}
where $f_{i\sigma}^{\dagger}$ is the creation operator for a fermion
carrying the spin of the electron, $b_{i}$ is the annihilation operation
for a boson carrying the charge. The no double occupancy constraint
$n_{i}\le1$ is replaced by an equality constraint 
\begin{equation}
\sum_{\sigma}f_{i\sigma}^{\dagger}f_{i\sigma}+b_{i}^{\dagger}b_{i}=1,\label{eq:conssb}
\end{equation}
which is enforced using a Lagrange multiplier \cite{Kotliar88a}.
In this formalism, $\sum_{i\sigma}f_{i\sigma}^{\dagger}f_{i\sigma}$
gives the total fermion density and $\sum_{i}b_{i}^{\dagger}b_{i}$
gives the total holon density. The invariance under the charge conjugation,
$f_{i\sigma}^{\dagger}\rightarrow e^{i\theta_{i}}f_{i\sigma}^{\dagger}$
and $b_{i}\rightarrow e^{-i\theta_{i}}b_{i}$ is implemented through
the constraint in Eq.~\ref{eq:conssb}. Other ways of fractionalizing
the electrons of course exist, with the celebrated $Z_{2}$ gauge
theory leading to ``visons'' excitations \cite{Senthil:2000eb,Senthil:2001jm},
or with extensions to higher groups like the pseudo-spin group SU(2)\cite{LeeWen97,Lee:1998cr}
or the spin group SU(2) \cite{Sachdev19,Ferraz11,Ivantsov18}.

\begin{figure}
\subfloat[]{\includegraphics[width=7cm]{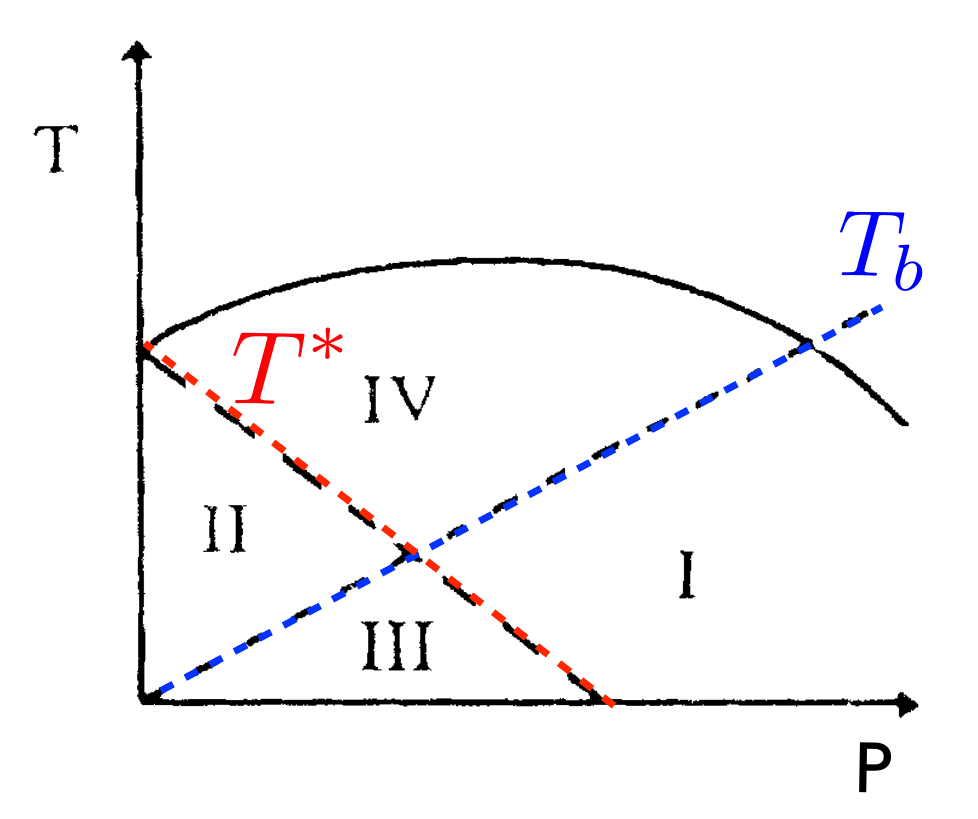}\label{Fig3a}

}

\subfloat[]{\includegraphics[width=7cm]{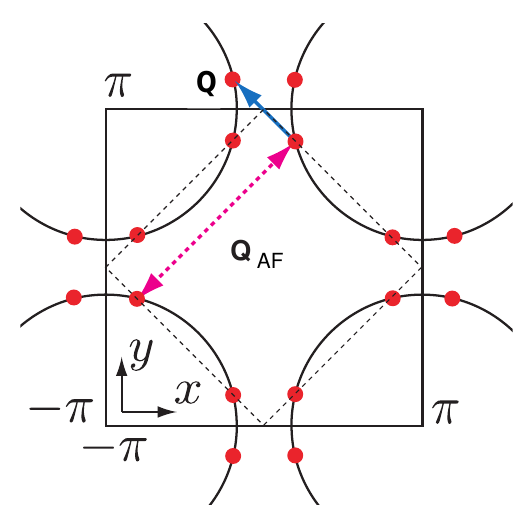}\label{Fig3b}

}

\subfloat[]{\includegraphics[width=5cm]{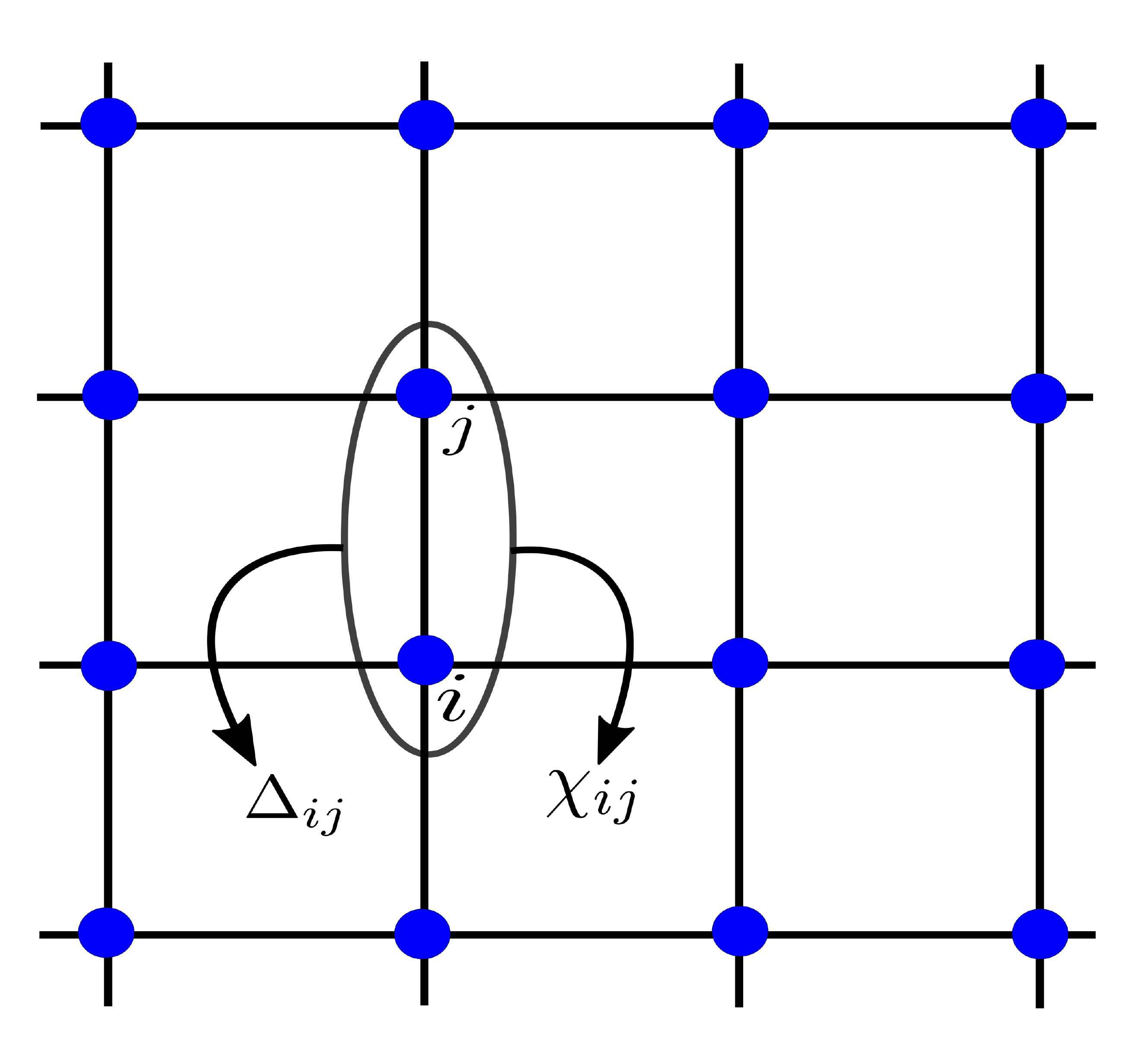}\label{Fig3c}

}

\caption{a) The phase diagram from strong coupling theories. The first line,
below which the Bose condensations (``holons'') occurs, and the
second line, below which sees the formation of ``spinon-spinon''
pairing, cross each other resulting into formation of a spin liquid
or ``Resonating Valence Bond'' (RVB) state. b) Schematic picture
of the eight hot spots (EHS) model. The dots in red are the eight
hot spots. The diagonal modulation wave vector is shown in blue (from
Ref. \cite{Kloss:2016hu}) c) Schematic depiction of bond variables
on a square lattice.}
\end{figure}

In all these cases the fractionalization of the electron into various
entities is effective in the ANR of the Brillouin zone, opening
a PG through the formation of small hole pockets. In this framework,
the global understanding of the phase diagram (as a function of doping)
goes in the following way (see Fig.~\ref{Fig3a}). A first line increasing
with doping, describes the Bose condensation of the holons. Implicit
is the assumption that the system is three dimensional so that the
Bose condensation $T_{b}$ occurs at finite temperature. A second
line $T^{*}$ decreasing with doping describes the PG phase and is
typically associated to a precursor of pairing. For example in the
U(1) theory, the $T^{*}$ line corresponds to the condensation of
spinon pairs $\left\langle f_{i\sigma}^{\dagger}f_{j\sigma}^{\dagger}\right\rangle $
on a bond. Strong coupling theories of the PG with electron fractionalization
can be considered as various examples of the celebrated Resonant Valence
Bond (RVB) theory \cite{Kotliar88b} introduced in the early days
after the discovery of cuprates, where some sort of spin liquid associated
with fluctuating bond states is considered as the key ingredient for
the formation of the PG \cite{Anderson87,Anderson04}. When the two
lines cross, below $T^{*}$and $T_{b}$ the electron re-confines so
that the SC transition is also a confining transition from the gauge
theory perspective. Above is the strange metal (SM) phase (Region
IV) whereas on the right we find the typical metallic- Landau Fermi
liquid phase (Region I).

Although this line of approach has been supported by a huge body of
theoretical work, we are skeptical that this is the final solution
for the PG phase. First, fractionalization of the electron produces,
when it happens, very spectacular effects like the quantization of
the resistivity, observed for example in the theory of fractional
Quantum Hall Effect (QHE). Here no spectroscopic signature of ``reconfinement''
of the electron has been observed at finite energy. Moreover, in most
of the under-doped region of cuprates, the ground state is a superconductor.
If the electron fractionalize at $T^{*}$ and that the ground state
is a superconductor, it means that at $T_{c}$ the electron shall
re-confine, form the Cooper pairs and globally freeze the phase of
the Cooper pairs. Of course it is possible but quite unlikely. A key
experiment which illustrates this feature is maybe ARPES measurements
which shows the presence of Bogoliubov quasi-particle in the ANR of
the Brillouin zone, in the under-doped region inside the SC phase
(see e.g. \cite{Vishik:2010fn}). Taken at its face value this experience
clearly hints at the formation of the SC state on the \textit{\textcolor{black}{whole
}}\textcolor{black}{Fermi surface, including the ANR, rather than
on small hole pockets centered around the nodes (see Fig.\ref{Fig1b}).
In this paper, we will now focus on the two other avenues of investigations
which are the fluctuations and the hidden phase transition at $T^{*}$.}

\section{Phase fluctuations}

The realization that phase fluctuations are important in the under-doped
region of the cuprates stems back from the seminal paper by Emery
and Kivelson, which pointed out that close to a Mott transition, the
localization of the electrons induces strong phase fluctuations of
all the fields present in the system \cite{EmeryVJ:1995dr}. The origin
of these fluctuations is deep and comes from the Heisenberg uncertainty
principle, in which the particle/wave duality ensures that the localization
of particles in space will generate phase fluctuations. Careful study
of the penetration depth as a function of doping shows that cuprates
are in the class of superconductors where the phase of the Cooper
pairs strongly fluctuates at $T_{c}$ leading to a Berezinsky Kosterlitz
Thouless transition typical of strong 2D fluctuations (see e.g. \cite{Benfatto:2000gy,Benfatto:2007df}).
Early experiments showed a linear dependence between the $T_{c}$
and the penetration depth, giving a lot of impetus to the fluctuations
scenario \cite{Uemura89,Homes04}.

Another phenomenology which can be successfully explained by phase
fluctuations of the Cooper pairs, lies in spectroscopic studies of
the spectral gap associated to the transition at $T^{*}$. Surface
spectroscopies like STM and ARPES tell us that the magnitude of the
spectral gap associated to the PG, which lies in the ANR of the Brillouin
zone, is unchanged when one goes through the SC $T_{c}$ \cite{Norman:1998hk,NormanKanigel07},
whereas the gap broadens more and more until $T^{*}$. On the other
hand, in the nodal region, the spectral gap vanishes at $T_{c}$ as
it should within the standard BCS theory. Moreover, we learn from
Raman scattering spectroscopies that in the $B_{1g}$ channel, which
scans the ANR, the spectral gap follows the $T^{*}$ line as a function
of doping, but is at the same time associated to the ``pair breaking''
below $T_{c}$ \cite{Loret19}.

These considerations lead to a very strong phenomenology based on
the phase fluctuations of the Coopers pairs, opening a channel for
damping between $T_{c}$ and $T^{*}$ and giving a remarkable
explanation for the filling of the AN spectral gap with temperature,
up to $T^{*}$ \cite{Norman:1995dd,NormanKanigel07,Banerjee:2011bz,Banerjee:2011cu,Boyack:2014fl,Boyack:2017gb,Chien2009}.
This theory was corroborated by the observation of an enormous Nernst
effect going up to $T^{*}$ in the Lanthanum and Bismuth compounds
which was interpreted as the presence of vortices up to a very high
temperature reaching very close to $T^{*}$ \cite{Li:2010gj,Li:2011cs,Li:2013ed}.

A long controversy followed, in order to determine the extend of the
temperature regime of phase fluctuations. Further experimental investigations,
including Nernst effect on YBCO compounds \cite{CyrChoiniere:2018ed}-where,
contrarily to the LSCO compounds, the contribution of the quasi-particles
has opposite sign from the contribution of the vortices, transport
studies \cite{RullierAlbenque:2011ji} and direct probing with Josephson
SQUID experiments \cite{Bergeal:2008gf}, led to the conclusion that,
the phase fluctuations extend only a few tens of degrees above $T_{c}$
but do not reach up to $T^{*}$. This issue can be resolved if we consider
a competition of the SC phase fluctuations with fluctuations of a
partner-competitor like particle-hole pairs, then it has been shown,
through the study of a non linear $\sigma$-model (NL$\sigma$M) \cite{Wachtel:2014ke},
that the true region of the SC phase fluctuations is reduced to a
temperature window close to $T_{c}$. In the same line of thoughts,
we examine here the possibility of extending fluctuations to a bigger
symmetry group, where not only the phase of the Cooper pairs fluctuates
but also there is a quantum superposition of Cooper pairs with a ``partner-competitor''
field.

\subsection{Extended symmetry groups and the case for SU(2) symmetry}

The idea to ``rotate'' the d-wave superconductor to another `partner',
hence enlarging the group allowed for fluctuations is present in many
theories for cuprates. A quick glance at the phase diagram of the
cuprates would convince anyone that the most prominent features are
the AF phase and the SC phase. Hence as a natural first guess, the
rotation from the SC to the AF state has been tried, leading to an
enlarged group of SO(5) symmetry \cite{Demler95,Demler04,Zhang97}.
The ten generators of the SO(5) group correspond to transitions between
the various states inside the quintuplet (three magnetic states, and
two SC states which are complex conjugate of one another). These theories
are based on an underlying principle that the `partners' are nearly
degenerate in energy. The `partners' posses an exact symmetry in the
ground state in some nearby parameter regime, for example the SC and
the AF states show exact degeneracy at half filling. At other parameters,
the symmetry is realized only approximately in the ground state. The
`hidden' exact symmetry emerges when the system is perturbed from
the ground state by increasing the temperature or the magnetic field.
In the framework of emergent symmetries, the PG is created by fluctuations
over a wide range of temperature. Later on, theories based on a SU(2)
gauge structure in the pseudo spin space led the formation of flux
phase \cite{Lee98}, and rotation from the SC state to flux phases
was envisioned. This theory had an emergent SU(2) symmetry with fluctuations
acting as well up to temperatures $T^{*}$.

Recently, new developments have shown observation of charge modulations
in the underdoped region of cuprate superconductors, in most of the
compounds through various probes, starting with STM \cite{Wise08,Hoffman02},
NMR \cite{Wu14,Wu11} and X-ray scattering \cite{Blackburn13a,Blanco-Canosa13,Blanco-Canosa14,Ghiringhelli12}.
At high applied magnetic field, the charge modulations reconstruct
the Fermi surface, forming small electron pockets \cite{Doiron-Leyraud07,LeBoeuf07,Sebastian12,Tabis14,Barisic2013,Grissonnanche:2015tl,Laliberte11,Chang2010}.
In YBCO, the charge modulations are stabilized as long range uniaxial
$\mathbf{Q}=\left(Q_{0},0\right)$ 3D order above a certain magnetic
threshold \cite{Gerber:2015gx,Chang:2016gz,Jang16} and the thermodynamic
lines can be determined with ultra-sound experiment \cite{LeBoeuf13,Laliberte2018}.
With the ubiquitous observation of charge modulations inside the PG
phase, a rotation of the SC state towards a Charge Density Wave (CDW)
state has been proposed \cite{Efetov13,Pepin14,Metlitski10b,Atkinson15,Freire:2015kg}.
This rotation has also two copies of an SU(2) symmetry group where
one rotates the SC state to the real $\Delta_{0}^{a}$ and imaginary
$\Delta_{0}^{b}$ parts of the particle-hole pair where $\left(i,j\right)$
are sites on a bond (see Fig.\ref{Fig3c}) with $\mathbf{r}_{j}=\mathbf{r}_{i}\pm a_{x,y}$
and $\mathbf{r}=\left(\mathbf{r}_{i}+\mathbf{r}_{j}\right)/2$ and
$\mathbf{Q}$ is the modulation wave vector corresponding to the CDW
state. In the case of the eight hot spots (EHS) model (defined after
Eq.~\ref{eq:3b}), $\mathbf{Q}$ is the the diagonal wave vector
$\mathbf{Q}=\left(Q_{0},Q_{0}\right)$ (shown in Fig.\ref{Fig3b})
but experimentally, it is axial with $\mathbf{Q}=\left(0,Q_{0}\right)$
or $\mathbf{Q}=\left(Q_{0},0\right)$. We have the two $l=1$ representations
$(\Delta_{-1},\Delta_{0},\Delta_{1})$ with 
\begin{align}
\Delta_{1} & =\frac{-1}{\sqrt{2}}\hat{d}\sum_{\sigma}\sigma c_{i\sigma}^{\dagger}c_{j-\sigma}^{\dagger}e^{i\left(\theta i+\theta_{j}\right)},\nonumber \\
\Delta_{0}^{a} & =\frac{1}{2}\hat{d}\sum_{\sigma}\left[c_{i\sigma}^{\dagger}c_{j\sigma}e^{iQ\cdot r+i\left(\theta i-\theta_{j}\right)}+c_{j\sigma}^{\dagger}c_{i \sigma}e^{-iQ\cdot r-i\left(\theta i-\theta_{j}\right)}\right],\nonumber \\
\nonumber \\
\Delta_{0}^{b} & =\frac{i}{2}\hat{d}\sum_{\sigma}\left[-c_{i\sigma}^{\dagger}c_{j\sigma}e^{iQ\cdot r+i\left(\theta i-\theta_{j}\right)}+c_{j\sigma}^{\dagger}c_{i \sigma}e^{-iQ\cdot r-i\left(\theta i-\theta_{j}\right)}\right],\nonumber \\
\nonumber \\
\Delta_{-1} & =\frac{1}{\sqrt{2}}\hat{d}\sum_{\sigma}\sigma c_{j-\sigma}c_{i\sigma}e^{-i\left(\theta i+\theta_{j}\right)}\label{eq:3}
\end{align}
and the corresponding $\eta$ - operators satisfy the SU(2) algebra,
\begin{align}
\left[\eta_{\pm},\Delta_{m}\right] & =\sqrt{l\left(l+1\right)-m\left(m\pm1\right)}\Delta_{m\pm1},\nonumber \\
\left[\eta_{z},\Delta_{m}\right] & =m\Delta_{m},\label{eq:3a}
\end{align}
with (note that $\eta_{z}$ is identical for both representations
$a$ and $b$) 
\begin{align}
\eta_{+}^{a} & =\frac{1}{2}\sum_{\sigma}\sigma[c_{i\sigma}^{\dagger}c_{i-\sigma}^{\dagger}e^{iQ\cdot r}e^{2i\theta_{i}}+c_{j\sigma}^{\dagger}c_{j-\sigma}^{\dagger}e^{-iQ\cdot r}e^{2i\theta_{j}}],\nonumber \\
\eta_{-}^{a} & =\frac{1}{2}\sum_{\sigma}\sigma[c_{i-\sigma}c_{i\sigma}e^{-iQ\cdot r}e^{-2i\theta_{i}}+c_{j-\sigma}c_{j\sigma}e^{iQ\cdot r}e^{-2i\theta_{j}}],\nonumber \\
\eta_{+}^{b} & =\frac{i}{2}\sum_{\sigma}\sigma[-c_{i\sigma}^{\dagger}c_{i-\sigma}^{\dagger}e^{iQ\cdot r}e^{2i\theta_{i}}+c_{j\sigma}^{\dagger}c_{j-\sigma}^{\dagger}e^{-iQ\cdot r}e^{2i\theta_{j}}],\nonumber \\
\eta_{-}^{b} & =\frac{-i}{2}\sum_{\sigma}\sigma[c_{i-\sigma}c_{i\sigma}e^{-iQ\cdot r}e^{-2i\theta_{i}}-c_{j-\sigma}c_{j\sigma}e^{iQ\cdot r}e^{-2i\theta_{j}}],\nonumber \\
\eta_{z} & =\frac{1}{2}\left[\eta_{+},\eta_{-}\right]=\frac{1}{2}\sum_{\sigma}\left(\hat{n}_{i\sigma}+\hat{n}_{j\sigma}-1\right).\label{eq:3b}
\end{align}

An exact realization of the emergent symmetry has been found, where
the Fermi surface is reduced to eight ``hot spots'' (crossing of the Fermi surface
and the AF zone boundary as shown in Fig.\ref{Fig3b}) and the electrons
interact with an AF critical mode in $d=2$ \cite{Efetov13,Pepin14,Metlitski10b,abanov03}.
In this simple model, the gap equations could be solved showing the
exact SU(2) symmetry between the Cooper pairing and particle-hole
channels. We observed some ordering of a composite order parameter,
which is a superposition of gaps in the particle-particle and particle-hole
channels 
\begin{align}
\hat{\Delta}^{*} \ket{0} & =\sum_{k\sigma}\left(c_{k\sigma}^{\dagger}c_{-k-\sigma}^{\dagger}+c_{k\sigma}^{\dagger}c_{k+Q\sigma}\right)\ket{0},\label{eq:4}
\end{align}
where the modulations in the charge sector occur at a finite diagonal
wave vector ${\bf Q}=\left(Q_{x},Q_{y}\right)$, where $Q_{x}$ and
$Q_{y}$ are the distance between two hot spots, as pictured in\textcolor{black}{{}
Fig.\ref{Fig3b}. A composite gap is formed at $T^{*}$ with the mean
square of the gaps in each channel 
\begin{align}
E^{*} & =\sqrt{\left|\chi\right|^{2}+\left|\Delta\right|^{2}},\label{eq:5}
\end{align}
} where $\chi$ is the gap in the particle-hole channel whereas $\Delta$
is the gap in the particle-particle channel.

Below $T^{*}$, a great amount of fluctuations are present which are
described by the O(4) NL$\sigma$M \cite{Efetov13,Einenkel14,Hayward14}.
For the four-fields $n_{\alpha}$,$\alpha=1\dots4$, with $n_{1}=\left(\Delta_{-1}+\Delta_{1}\right)/2$,
$n_{2}=\left(\Delta_{-1}-\Delta_{1}\right)/2$, $n_{3}=\Delta_{0}^{a}$
, $n_{4}=\Delta_{0}^{b}$ the effective action writes 
\begin{align}
S & =1/2\int d^{d}x\sum_{\alpha=1}^{4}\left(\partial_{\mu}n_{\alpha}\right)^{2}, & \mbox{with } & \sum_{\alpha=1}^{4}\left|n_{\alpha}\right|^{2}=1.\label{eq:6}
\end{align}

\subsection{What can the model explain at this stage?}

\subsubsection{Charge modulations inside the vortex core}

\begin{figure}
\subfloat[]{\includegraphics[width=7cm]{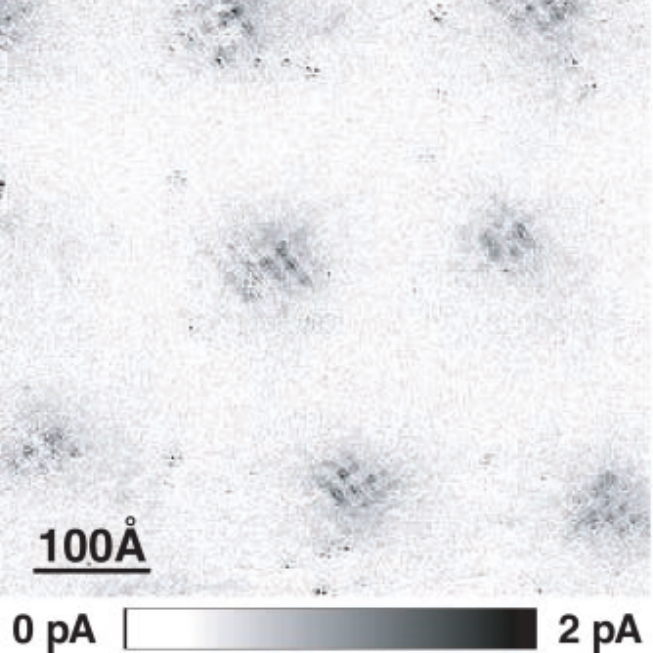}\label{Fig4a}

}

\subfloat[]{\includegraphics[width=7cm]{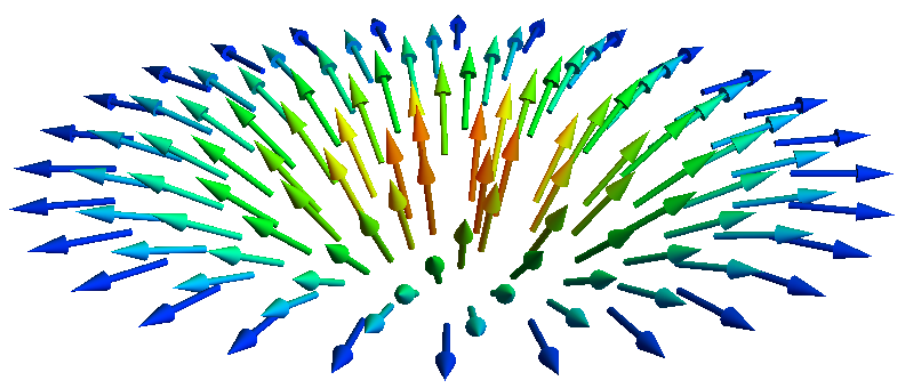}\label{Fig4b}

}

\caption{a) Charge modulations observed in the vortex core, from the seminar
paper \cite{Hoffman02}. The data were taken at magnetic field $B=0T$
and $B=8T$ and then subtracted. Charge modulations ``appreared''
inside the vortex core. b) Schematic picture of a Skyrmion in the
$\eta$-space corresponding to the O(3) NL$\sigma$M in Eq.(\ref{eq:12}).
The axes are $\left(m_{x},m_{y},m_{z}\right)$.}
\end{figure}

At this stage, the model can already explain a few properties of cuprate
superconductors. Since the model treats very seriously the competition
between CDW and SC pairing order parameters, it predicts charge modulations
inside the vortex\textcolor{black}{{} core (Fig.\ref{Fig4a})\cite{Hoffman02,Wise08,Hamidian15a,Wu13a,Wu14}.
Indeed, since the SC order parameter vanishes there,} \textcolor{black}{the
competing order emerges at the core. The special structure associated
to this feature is called a meron, or half skyrmion, in the pseudo-spin
space. It can be noted that it is a generic prediction of the theories
of emergent symmetries, that the competing order shows up inside the
vortex core. For example, the SO(5) theory predicts AF correlations
inside the vortex core \cite{Arovas97,Ghosal02}, whereas the SU(2)
symmetry which rotates superconductivity to the $\pi$-flux phase
predicts that the $\pi$-flux orbital state \cite{Lee:2001ib,Lee06}
is present inside the vortex core. Although AF correlations were observed
in the vortex core in the La-compounds \cite{Lake01}, for YBCO, BSCCO
and Hg-series, STM experiments \cite{Hoffman02,Wise08} and NMR \cite{Wu13a}
gave evidence for charge modulations. This ubiquitous observation
of charge modulations inside the vortex core is a nice test for the
theory of emergent SU(2) symmetry, but the presence of charge modulations
inside the vortex core could also be explained by a strong competition
between SC and CDW without invoking any emergent symmetry \cite{Kivelson:2002er,Zhang:2002hz}.}

\subsubsection{B-T phase diagram}

\begin{figure}
\subfloat[]{\includegraphics[width=7cm]{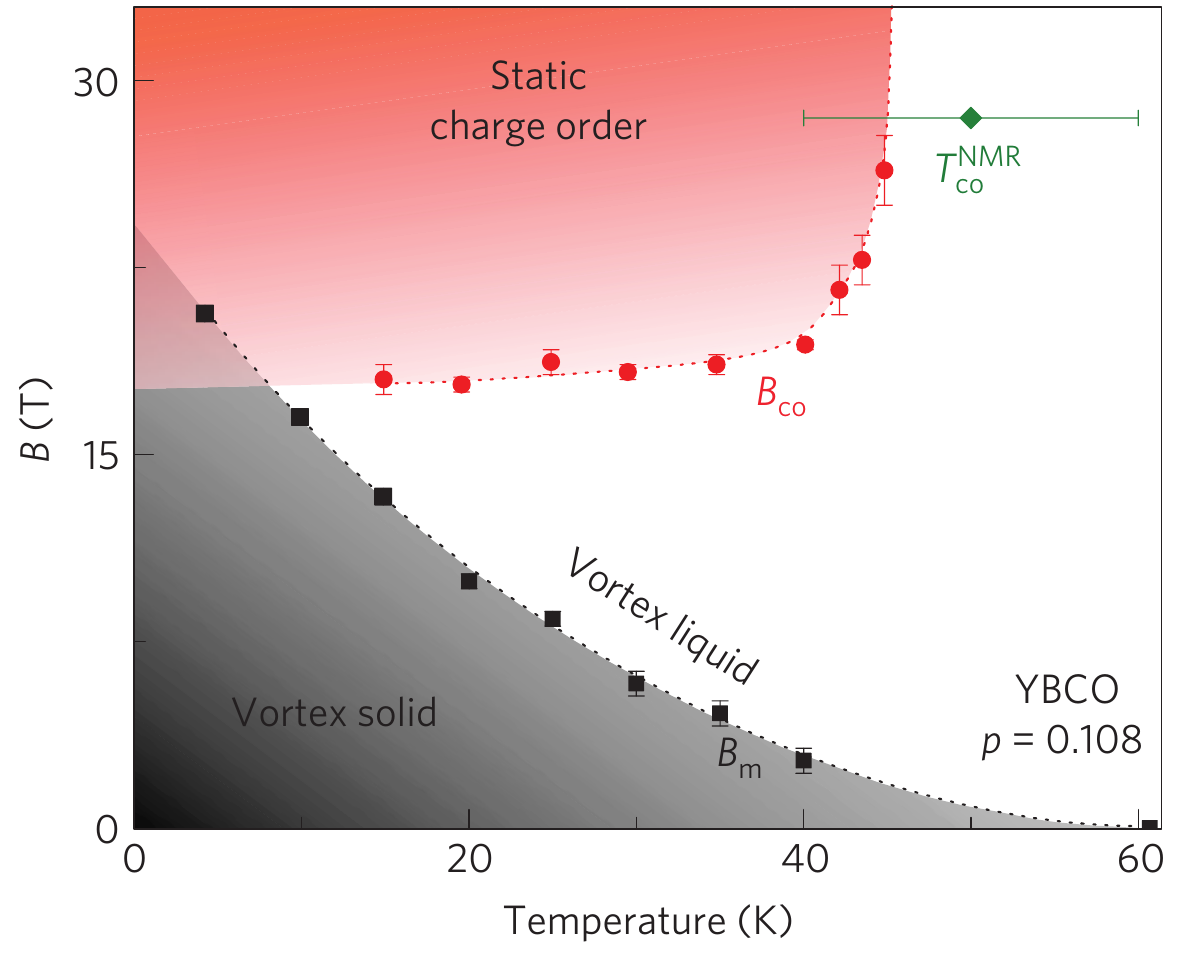}\label{Fig5a}

}

\subfloat[]{\includegraphics[width=7cm]{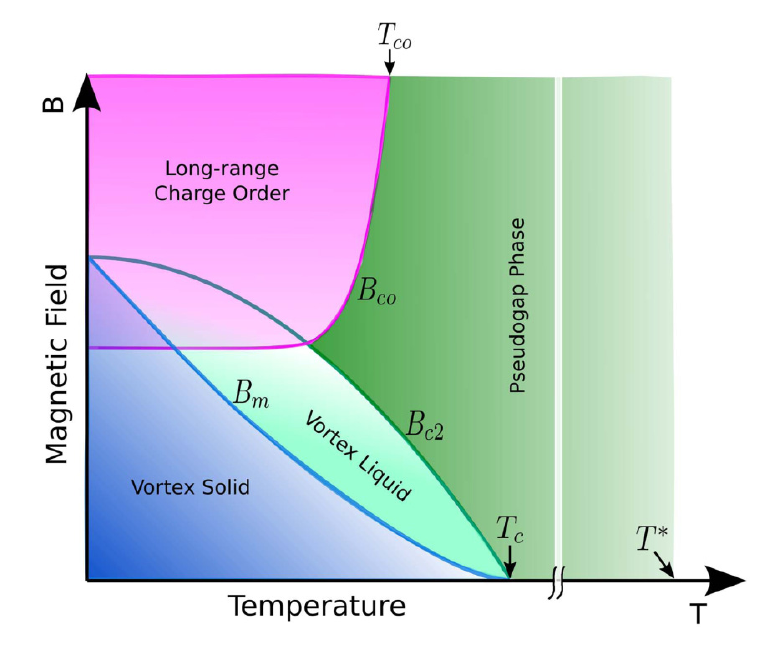}\label{Fig5b}

}

\caption{a) Experimental $B-T$ phase diagram from Ref.\cite{LeBoeuf13}. The
shape of the lines is done with ultrasound experiments. b) Schematic
phase diagram described through the SU(2) theory for competition between
charge modulations and Cooper pairing \cite{Chakraborty18}.}
\end{figure}

This leads us to a second set of experimental evidence, explicitly
the phase diagram in the presence of an applied magnetic field as
described in Figs.\ref{Fig5a}, \ref{Fig5b}\cite{LeBoeuf13,Chakraborty18}.
For the compound YBCO, a phase diagram could be derived as a function
of an applied magnetic field up to roughly $20T$. For this specific
compound, one observes at $H_{c}=17T$ \cite{LeBoeuf13}, a second
order phase transition towards a 3D charge order (CO) state with one
uniaxial vector of modulations \cite{Gerber:2015gx,Chang16}. The
shape of this transition is very flat in temperature \cite{LeBoeuf13,Laliberte2018},
a fact that cannot be accounted by a simple model of competition between
the two orders, but can be explained by a pseudo-spin flop transition,
where the system suddenly goes from the SC state to the CO. The model
of pseudo-spin flop comes directly from the expression of the NL$\sigma$M
where the constraint plays the role of the value of the spin in a
magnetic spin-flop transition. The pseudo-spin flop transition accounts
for the flatness \cite{Meier13} in temperature of the transition,
but it does not explain the phase of co-existence \cite{LeBoeuf13,Wu13a,Kacmarcik18}.
Accounting for this phase requires to break the exact SU(2) symmetry
underlying the NL$\sigma$M by an amount of roughly 5\%. As a conclusion,
an SU(2) emergent symmetry and its NL$\sigma$M can account for the
whole phase diagram owing to the fact that it is broken by 5\% \cite{Chakraborty18}.

\subsubsection{Collective modes}

An emergent symmetry is characterized by a set of collective modes\textcolor{black}{{}
\cite{Morice18b}}. Let us assume that the CDW is real, as in the first
set of Eq.\eqref{eq:3}. Here the operators $\eta$ and $\eta^{*}$
which enable the ``rotation'' from the SC to the CDW states are
the generators of an O(3) Lie algebra 
\begin{align}
{\cal L}= & \left(\begin{array}{ccc}
0 & * & *\\
-\frac{i}{2}\left(\eta_{+}^{a}-\eta_{-}^{a}\right) & 0 & *\\
-\eta_{z} & \frac{1}{2}\left(\eta_{+}^{a}+\eta_{-}^{a}\right) & 0
\end{array}\right),\label{eq:7}
\end{align}
where only the operator $\eta^{a}$ in Eq.(\ref{eq:3b}) contribute
(since we rotate to the real part of the particle-hole pair), and
where the notation $*$ stands for a hermitian matrix. The structure
of ${\cal L}$ in Eq.(\ref{eq:7}) give rise to collective modes.
These collective modes are spin zero, charge two, and reflect the
structure of the SU(2) symmetry. They can be considered as Pair Density
Wave (PDW) excitations since they have non zero center of mass wave
vector. They could be responsible for the mode observed in the $A_{1g}$
channel in Raman Scattering \cite{Benhabib:2015ds}\textcolor{black}{{}
and can also be seen by spectroscopy experiments, like X ray, MEELS
\cite{Vig:2017da,Mitrano:2018dj} or soft X-rays \cite{Chaix:2017fs},
where the resolution in $\mathbf{q}$-space can be traced. The theory
predicts that the mode occurs around the same wave vector as the charge
modulations and has a typical linear shape. Using the EHS model, we
could fit the slope of the mode to a recent X-ray experiments on BSCCO
compounds \cite{Chaix:2017fs}. }

\subsubsection{Symmetry of the spectral gap with respect to zero energy}

The PG state is identified in various spectroscopies with either the pair-breaking peak as seen in the $B_{1g}$ channel in Raman \cite{Sacuto:2013ez,Devereaux07}, or the gap seen in STM \cite{Kugler01} or ARPES \cite{Vishik:2010fn,Damascelli03,Vishik18}. The spectral ``peak'' or the gap concern the ANR of the Brillouin zone. One most noticeable feature about this gap is that it goes ``unchanged'' when one decreases
the temperature to reach inside the SC phase \cite{Vishik:2010fn,Damascelli03,Vishik18}.
Said in other words, the SC gap in the ANR below $T_{c}$ remains
the same when the temperature is raised up to $T^{*}$. This very
unusual feature contrasts with the spectroscopic behavior around the
node, as seen for example in the $B_{2g}$ channel in Raman spectroscopy
\cite{Sacuto:2013ez,Devereaux07}, or in ARPES resolved in \textbf{$\text{k}$}-space
\cite{Vishik18}, where the gap as a function of temperature vanishes
at $T_{c}$ which is typical of a standard BCS behavior. The behavior
in the ANR naturally calls for an interpretation in terms of
the phase fluctuations of the SC pairing order parameter whereas the
phase coherence sets up below $T_{c}$. This led to the development
of a powerful phenomenology of the PG \cite{Lee14,Montiel:2016it},
in which one distinguished feature is that the spectral gap is symmetric
around $E=0$ \cite{Verret:2017he}. This feature is easily reproduced
by any theory of preformed pairs and emergent symmetries. On the other
hand, all other theories will produce an asymmetry of the PG around
$E=0$ (see e.g. Ref. {[}\onlinecite{Wu18,Sakai16,Sakai:2018jq}{]}).

A somewhat related issue is the evolution
of the number of carriers with doping, studied in a recent Hall measurement
\cite{Badoux16}. It is shown that the number of carriers evolves
first linearly as the doping $p$, and then goes quite abruptly to
$1+p$, forming a large hole Fermi surface around the critical doping
$p^{*}$. This behavior can be accounted by all the theories which
open a gap around the eight hot spots (see Fig.\ref{Fig3b}), including theories
coming from strong coupling scenarios \cite{Yang:2006eq,Chatterjee17,Storey16,Verret:2017he} or theories
which open a gap on weaker coupling scenarios with standard symmetry breaking \cite{Morice:2017kd,Storey16,Verret:2017he}.

\subsection{Microscopic models for emergent symmetries}

A first generalization of the EHS model, was to start with a model
for short range AF correlations, rather than a model of an AF QCP
with critical fluctuations in 2D. A possible model consists of a simplification
of the t-J model of Eq.(\ref{eq:1}), where the constraint of no double
occupancy in the Gutzwiller projection operators \cite{Parmekanti01}
is treated at the mean-field level and where an extra nearest neighbor
(n.n.) Coulomb interaction is retained 
\begin{align}
H & =\sum_{i,j,\sigma}{t_{ij}\ (c_{i,\sigma}^{\dagger}c_{j,\sigma}+h.c)}\nonumber \\
 & +\sum_{i,j}{J_{i,j}\ \bm{S}_{i}\cdot\bm{S}_{j}+V_{i,j}\ n_{i}n_{j}},\label{eq:8}
\end{align}
where $c_{i,\sigma}^{\dagger}$ ($c_{i,\sigma}$) is a creation (annihilation)
operator for an electron at site $i$ with spin $\sigma$, $n_{i}=\sum_{\sigma}c_{i,\sigma}^{\dagger}c_{i,\sigma}$
is the number operator and $\bm{S}_{i}=c_{i,\alpha}^{\dagger}\bm{\sigma}_{\alpha,\beta}c_{i,\beta}$
is the spin operator at site $i$ ($\bm{\sigma}$ is the vector of
Pauli matrices). $J_{i,j}$ is an effective AF coupling which comes
for example from the Anderson super-exchange mechanism \cite{Montiel16}.
\textcolor{black}{The constraint of no double occupancy typical of
the strong Coulomb onsite interaction is implemented through the Gutzwiller
approximation by renormalizing the hoping parameter and the spin-spin
interaction with $t\left(p\right)=g_{t}t=\frac{2p}{1+p}t$ and $J\left(p\right)=g_{J}J=\frac{4}{(1+p)^{2}}J$,
where $p$ is the hole doping while the density-density interaction
is left invariant. }We also assume that the AF correlations are dynamic,
strongly renormalized, and short ranged, as given by the phenomenology
of Neutron scattering studies for cuprates \cite{Hinkov07} and $V_{i,j}$
is a residual Coulomb interaction term.

The theory of the SU(2) emergent symmetry, although producing some
strong phenomenology suffers from a few weaknesses. Maybe the most
prominent one is the stability of the symmetry with respect to a small
change of parameters in the model \cite{Wang18a}. Let's consider
again the EHS model which has an exact realization of the symmetry.
If we try to generalize to a real Fermi surface, we notice that the
symmetry is valid only at the hot spots, but anywhere away in the
Brillouin zone it is broken. In order to resolve this issue, we first
enlarged the space of the charge modulations by considering multiple
wave vectors depending on the position in \textbf{k}-space. Although
a bit artificial, this trick was developed in order to maximize the
area in \textbf{k-}space where the SU(2) symmetry is realized. The
object thus created in the charge channel has the form of charge excitons
and forms droplets in real space. This enabled us to have a theory
for the proliferation of droplets in the under-doped region of cuprate
superconductors \cite{Montiel:2017gf,Montiel16}. Formation of droplets,
or phase separation in real space, is a very interesting idea to explain
the PG region because it has been observed through NMR experiments
that the PG phase is very robust with respect to induced disorder
\cite{Alloul14}. When a vacancy is induced in the compound, for example
through $Zn$ doping \cite{Alloul91}, or irradiation with electrons
\cite{RullierAlbenque:2000fl}, most of the energy scales get powerfully
reduced, like the SC $T_{c}$ or the charge ordering temperature $T_{co}$,
but the PG line remains unaffected. More precisely, NMR studies tell
us that around the $Zn$- impurity a small region of AF correlations
is forming, as if the close vicinity of the impurity was revealing
the underlying presence of short range AF correlations. But in-between the $Zn$- impurities, we recover the unaffected PG. This tells us
that the PG is made of some short range ``droplets'' ( in real space),
which don't ``communicate'' with one another, so that when a strong
local perturbation is present ( like the $Zn$- impurity) the PG is
only affected locally \cite{Alloul91,Alloul09}.

\begin{figure}
\includegraphics[width=7cm]{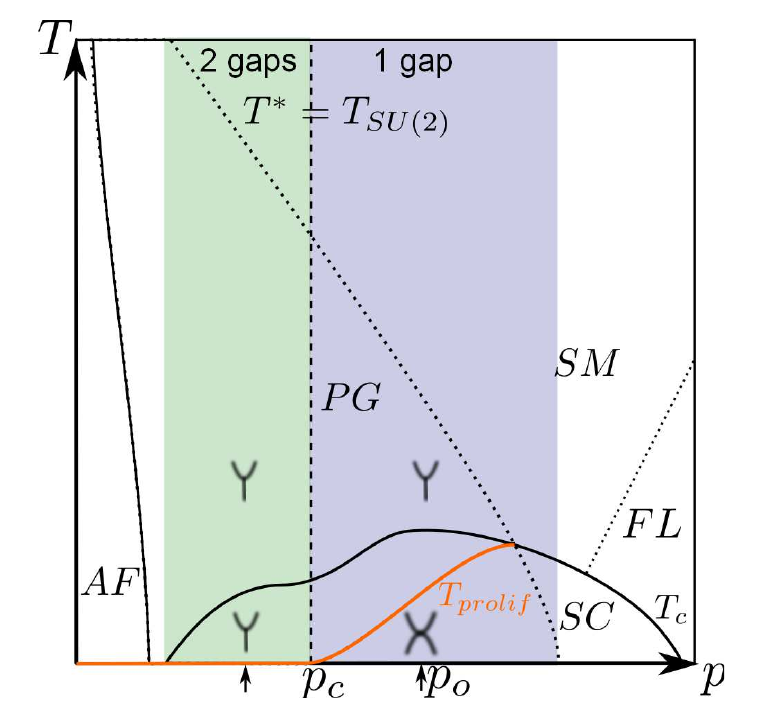}\
\caption{Phase diagram for the Inelastic Neutron Scattering response at $\left(\pi,\pi\right)$
from Ref.\cite{Montiel17}.}
\label{Fig6} 
\end{figure}

We identify two regimes as a function of doping (Fig.~\ref{Fig6})
\cite{Montiel17}. At lower doping in the under-doped region ($0.06<x<0.12)$
the droplets proliferate, which translates itself as a mixed character
of the PG in the AN region with superposition of charge and SC gaps.
This induces a special response in the Inelastic Neutron Scattering
(INS), with the spin one mode observed at the AF wave vector $\left(\pi,\pi\right)$
having a ``Y''shape typical of the PG phase. The INS mode is treated
in this theory as coming from spin excitons inside the SC phase or
in the PG phase. On the other hand, at larger doping ($0.12<x<0.20$),
the droplets become rarer and the gap in the ANR below $T_{c}$ is
a Cooper pair gap. In this region, the INS mode changes form below
$T_{c}$ with now an ``hourglass'' or typical ``X'' shape especially
visible in the YBCO compound, where the splitting between the lower
branches of the ``X'' coming from the border of the particle-hole
continuum inside the SC phase (see \cite{Hinkov07} and reference
therein).

The formulation of the SU(2) symmetry using multiple wave vectors
was intended to preserve the symmetry in the ANR of the Brillouin
zone. Although the possibility of multiple wave vectors is interesting,
it is not very likely that a physical system does not choose one wave
vector at least at lower temperature. Is there another route to get
an organizing principle to treat the competition between the charge
modulations and the SC without resorting to the multiple wave vectors
trick?

\section{Phase transition and Higgs Mechanism at $T^{*}$}

To summarize, in the last section, we gave an attempt of the generalization
of the SU(2) symmetry for realistic Fermi surfaces and more realistic
starting point of a model of short range AF interactions. The price
we have to pay is to consider multiple wave vectors for the charge
modulations. Although such a scenario is valid in theory, one could
ask whether it is effectively realized in the experiments and whether
we could get a more robust mechanism to get fluctuations at the $T^{*}$
line.

\subsection{Chiral model and Hopf mapping}

A new idea came by noticing that the SU(2) symmetry that rotates the
SC phase to the CDW modulations comes with two copies of SU(2) in
the case (like in the EHS model) where the charge sector is a ``bond
exciton''- with a real and imaginary part \cite{Chakraborty19}.
As in the O(4) NL$\sigma$M Eq.(\ref{eq:6}), we have two complex
operator fields 
\begin{align}
z_{1} & =\hat{d}\sum_{\sigma}\sigma c_{j-\sigma}c_{i\sigma}\equiv\Delta_{ij},\nonumber \\
z_{2} & =\hat{d}\sum_{\sigma}c_{i\sigma}^{\dagger}c_{j\sigma}e^{i\mathbf{Q}\cdot\left(\mathbf{r}_{i}+\mathbf{r}_{j}\right)/2}\equiv\chi_{ij},\label{eq:9}
\end{align}
with $\hat{d}$ being the d-wave form factor and the constraint writes 
\begin{align}
E^{*} & =\sqrt{\left|z_{1}\right|^{2}+\left|z_{2}\right|^{2}},\label{eq:10}
\end{align}
where $E^{*}$ is a high energy scale which is interpreted as the
PG energy scale in the case of the cuprate superconductors. Considering
the spinor field $\psi=\left(\begin{array}{c}
z_{1}\\
z_{2}
\end{array}\right)$, the constraint in Eq.(\ref{eq:10}) is invariant with respect to
factorizing a global phase $\psi\rightarrow e^{i\theta}\psi$. This
gauge invariance is visible in the SU(2) chiral model where the fields
are allowed to fluctuate within an SU(2) matrix. 
\begin{align}
S=\frac{1}{2}\int d^{d}x & Tr[\partial_{\mu}\varphi^{\dagger}\partial_{\mu}\varphi], & \mbox{ with } & \varphi_{ab}=\frac{\delta_{ab}}{2}-z_{a}z_{b}^{*},\nonumber \\
\mbox{ and } & \sum_{a=1}^{2}z_{a}^{*}z_{a}=1.\label{eq:11}
\end{align}
Using the Hopf mapping of the sphere $S_{3}$ represented by Eq. (\ref{eq:10})
to the sphere $S_{2}$, the model in Eq.(\ref{eq:11}) can be reduced
to and O(3) NL$\sigma$M as follows (see a generic proof in \cite{Fradkinbook}).
We define $m^{a}=z_{\alpha}^{*}\sigma_{\alpha\beta}^{a}z_{\beta}$
where $\sigma$ are Pauli matrices and the indices $a=x,y,z$, the
effective action is now 
\begin{align}
S & =1/2\int d^{d}x\sum_{a=1}^{3}\left(\partial_{\mu}m^{a}\right)^{2}, & \mbox{with } & \sum_{a=1}^{3}\left|m^{a}\right|^{2}=1.\label{eq:12}
\end{align}

We see that the form in Eq.(\ref{eq:12}) is similar to the one of
Eq.(\ref{eq:6}) but with three fields (O(3) NL$\sigma$M) rather
than four (O(4) NL$\sigma$M). In the case of the two fields of Eq.(\ref{eq:9})
which represent the case of cuprates, the operators $m_{x}$, $m_{y}$
and $m_{z}$ correspond to PDW fluctuations.

\subsection{$U(1)\times U(1)$ gauge theory and Higgs phase transition at $T^{*}$}

The two complex fields $\Delta_{ij}$ and $\chi_{ij}$ in Eq.~\eqref{eq:9}
are defined on bonds $(i,j)$ with $\mathbf{r}_{j}=\mathbf{r}_{i}\pm a_{x,y}$
(see Fig.\ref{Fig3c}). $\Delta_{ij}$ and $\chi_{ij}$ represent
preformed pairs in the particle-particle (PP) channel and the particle-hole
(PH) channel. In order to construct a continuum field theory, these
fields are defined on the midpoint of the bond, $\mathbf{r}=\left(\mathbf{r}_{i}+\mathbf{r}_{j}\right)/2$
such that $z_{1}(r)=\Delta_{ij}$ and $z_{2}(r)=\chi_{ij}$. The effective
field theory will then have a U(1) $\times$ U(1) gauge structure.
One U(1) corresponds to the usual charge symmetry (usually broken
by superconducting ground state) and the other is a consequence of
the fact that we have pairs on bonds. By writing $z_{1}$ and $z_{2}$
on bonds, we have doubled the gauge structure to U(1) $\times$ U(1)
with two gauge fields introduced in order to accommodate two independent
phases. Without any loss of generality, the U(1) $\times$ U(1) gauge
theory can be formulated in terms of a global and a relative phase
of the two kinds of preformed pairs. In terms of the spinor $\psi$,
the global and relative phases can be expressed as follows 
\begin{align}
\psi & =e^{i\theta}e^{i\tau_{3}\varphi}\left(\begin{array}{c}
\tilde{z}_{1}\\
\tilde{z}_{2}
\end{array}\right),\label{eq:13}
\end{align}
where $\tilde{z}_{1}$ and $\tilde{z}_{2}$ are two fields which can
carry d-wave symmetry and modulations. Typically here, $\tilde{z}_{1}=\hat{d}\left|z_{1}\right|$
and $\tilde{z}_{2}=\hat{d}\left|z_{2}\right|e^{i\mathbf{Q}\cdot\mathbf{r}}$. Two gauge fields $a_{\mu}$ and $b_{\mu}$ are introduced in the theory
to enforce the gauge invariance. The transformation $\psi\rightarrow e^{i\theta}e^{i\tau_{3}\varphi}\psi$,
$a_{\mu}\rightarrow a_{\mu}+\partial_{\mu}\theta$,
$b_{\mu}\rightarrow b_{\mu}+\partial_{\mu}\varphi$ leaves the following
action
{\small{
\begin{align}
{\cal S}_{a,b} & =\int d^{d}x\left[\frac{1}{2g}\left|D_{\mu}\psi\right|^{2}+V\left(\psi\right)+\frac{1}{4}F_{\mu\nu}F^{\mu\nu}+\frac{1}{4}\tilde{F}_{\mu\nu}\tilde{F}^{\mu\nu}\right],\label{eq:1}\\
\mbox{with } & D_{\mu}=\partial_{\mu}-ia_{\mu}-i\tau_{3}b_{\mu},\nonumber \\
 & F_{\mu\nu}=\partial_{\mu}a_{\nu}-\partial_{\nu}a_{\mu},\nonumber \\
\mbox{and } & \tilde{F}_{\mu\nu}=\partial_{\mu}b_{\nu}-\partial_{\nu}b_{\mu},\nonumber 
\end{align}}}
invariant, where $\tau_{3}$ is the Pauli matrix in the spinorial space $\psi$ and $a_{\mu}$ and $b_{\mu}$ are gauge fields corresponding respectively to the spinor's global phase $\theta$ and relative phase $\varphi$. A linear combination of the two gauge fields, $\mathbf{a}+\mathbf{b}=2\mathbf{A}$ gives twice the electro-magnetic gauge field and another combination, $\mathbf{a}-\mathbf{b}=\tilde{\mathbf{a}}$ is identified as a dipolar field on a bond.

The concept of two kinds of preformed pairs and the resultant U(1)
$\times$ U(1) gauge structure opens up unique possibilities with
hierarchy of phenomenon occurring with reduction in temperature from
a high value. At $T^{*}$, the system encounters a first Higgs mechanism
where the global U(1) phase $\theta$ of the spinor $\psi$ gets frozen.
As a result, one gauge field acquires a mass $E^{*}=\sqrt{\left|\chi_{ij}\right|^{2}+\left|\Delta_{ij}\right|^{2}}$
opening a gap in the ANR of the Brillouin zone and $E^{*}$ characterizes
the PG energy scale. Owing to this Higgs mechanism, the PG line will
show universal features independent of the material specifics or disorder.
Thus at $T^{*}$, the PP and PH pairs get entangled and compete strongly
with each other. Due to this competition, both the amplitudes $\left|z_{1}\right|$
and $\left|z_{2}\right|$ fluctuate wildly along with fluctuations
in the relative phase. The effective theory just below $T^{*}$ corresponds
to the O(3) NL$\sigma$M or, equivalently, to the SU(2) chiral model,
and the typical excitations are made of PDW $\eta$ -modes. The freezing
of the global phase at $T^{*}$ has removed one copy of the O(3) NL$\sigma$M
leaving the other one untouched. It is to be noted that the O(3) NL$\sigma$M
has topological excitations which are skyrmions in the pseudo-spin
space and may account for the recently observed anomalous thermal
Hall effect \cite{Grissonnanche19}. To explain the anomalous Hall effect, there are other recent proposals based on proximity to a quantum critical point of a `semion' topological ordered state \cite{Chatterjee19}, presence of spin-dependent next-nearest neighbor hopping in the $\pi$-flux phase \cite{Han19} or presence of large loops of currents \cite{Varma19}. If we continue to lower the temperature
in the under-doped region, different temperature lines emerge (see
Fig.~\ref{fig:The-phase-diagram}). First, there are two cross-over
mean-field lines corresponding to the condensation of the amplitudes
of each field $\Delta_{0}\left(T\right)=\left|z_{1}\right|_{0}$ and
$\chi_{0}\left(T\right)=\left|z_{2}\right|_{0}$, leading to a uniform
d-wave Cooper pairing contribution below $T_{fluc}$ and modulated
d-wave charge contribution below $T_{co}$. Below each line, the preformed
pairs have a non zero precursor gap in each channel, and these two
channels compete. The two fields $z_{1}$ and $z_{2}$ still satisfy
Eq. (\ref{eq:10}) with a condensed part ($\Delta_{0}$ and $\chi_{0}$)
and a fluctuation part such that $\left|z_{1}\right|=\Delta_{0}\left(T\right)+\delta\left|z_{1}\right|$
and $\left|z_{2}\right|=\chi_{0}\left(T\right)+\delta\left|z_{2}\right|$.
These lines do not have to be identical. In\textcolor{red}{{} }\textcolor{black}{Fig.}\textcolor{red}{\ref{fig:The-phase-diagram}
}\textcolor{black}{they are represented, with the first one $T_{co}$
corresponding to a quasi long range charge order (experimentally this
order is bi-dimensional, so that it is impossible to have it fully
long range at finite temperature) and the second one $T_{fluc}$ corresponding
to the phase fluctuation regime of the Cooper pairs. Note than when
$T_{fluc}<T_{co}$, a PDW composite order is induced below $T_{fluc}$,
since the field $z_{1}z_{2}=\Delta$$\chi$ has the symmetry of a
SC order with non-zero center of mass and its phase was frozen at
$T^{*}$. These lines are determined by solving the gap equations
in each channel, corresponding to the microscopic model in Eq.(\ref{eq:8}).
Below $T_{c}$, the second freezing of the relative phase occurs and
we are fully in the SC phase. One consequence of our theory, is that
the SC ground state is actually a super-solid (or PDW), a fact that
is supported by X-Ray experiments, which show that the charge order
signal does not go to zero as $T\rightarrow0$ in the SC phase (see
e.g. \cite{Blackburn13a}).} Since our theory of preformed pairs is
based on two complex fields, the phase in the particle-hole channel
produces currents of d-density wave (dDW) \cite{Chakravarty01} type,
as well as a series of pre-emptive transitions breaking $C_{4}$ symmetry
(nematicity), time reversal symmetry (TR) and creating loop currents.
The reasoning follows the steps of earlier works \cite{Wang14,Wang15a,Wang15b}.
\textcolor{black}{Another consequence of this double stage freezing
of the phase is that the phase slip associated to the charge ordering
is now stiff and related to the phase of the superconductor. This
very unusual situation promises to open space for future experimental
verifications.}

\subsection{What can this work explain?}

The main idea of this work is to have two sets of preformed pairs,
one in the particle-hole channel, that we called ``bond excitons'',
and one in the particle-particle channel-the Cooper pairs that become
entangled at $T^{*}$. The freezing of the entropy down to $T=0$
then follows its own route showing cascade of phenomenon occurring at different temperatures. The notion of preformed pairs shall be distinguished from the typical
scenario of phase fluctuations \cite{Banerjee:2011bz,Benfatto:2007df}
around a mean field amplitude. Indeed, in the phase fluctuations scenario,
the focus is put on the fluctuations only with no restrictions on
the size of the pairs, whereas the preformed pair scenarios \cite{Chien2009,Boyack:2014fl,Boyack:2017gb}
put the accent on the very short size of the Cooper pair which opens
a wide region in temperature where the pairs behave like a hard core
boson and undergo Bose Einstein condensation. In view of the strong
correlations in cuprates and the very short size of the Cooper pairs,
the concept of preformed pairs is very natural. Moreover, recent
experiments have revived the issue of fluctuating preformed pairs with their observation up to $T^*$ in pump probe measurements \cite{Rajasekaran18,Hu14,Fausti11} and in the over-doped region of the phase diagram in time domain spectroscopy \cite{Mahmood:2019gk}. We review below a few experiments which could be explained by such a scenario.

\subsubsection{A precursor in the charge channel}

\begin{figure}
\subfloat[]{\includegraphics[width=6cm]{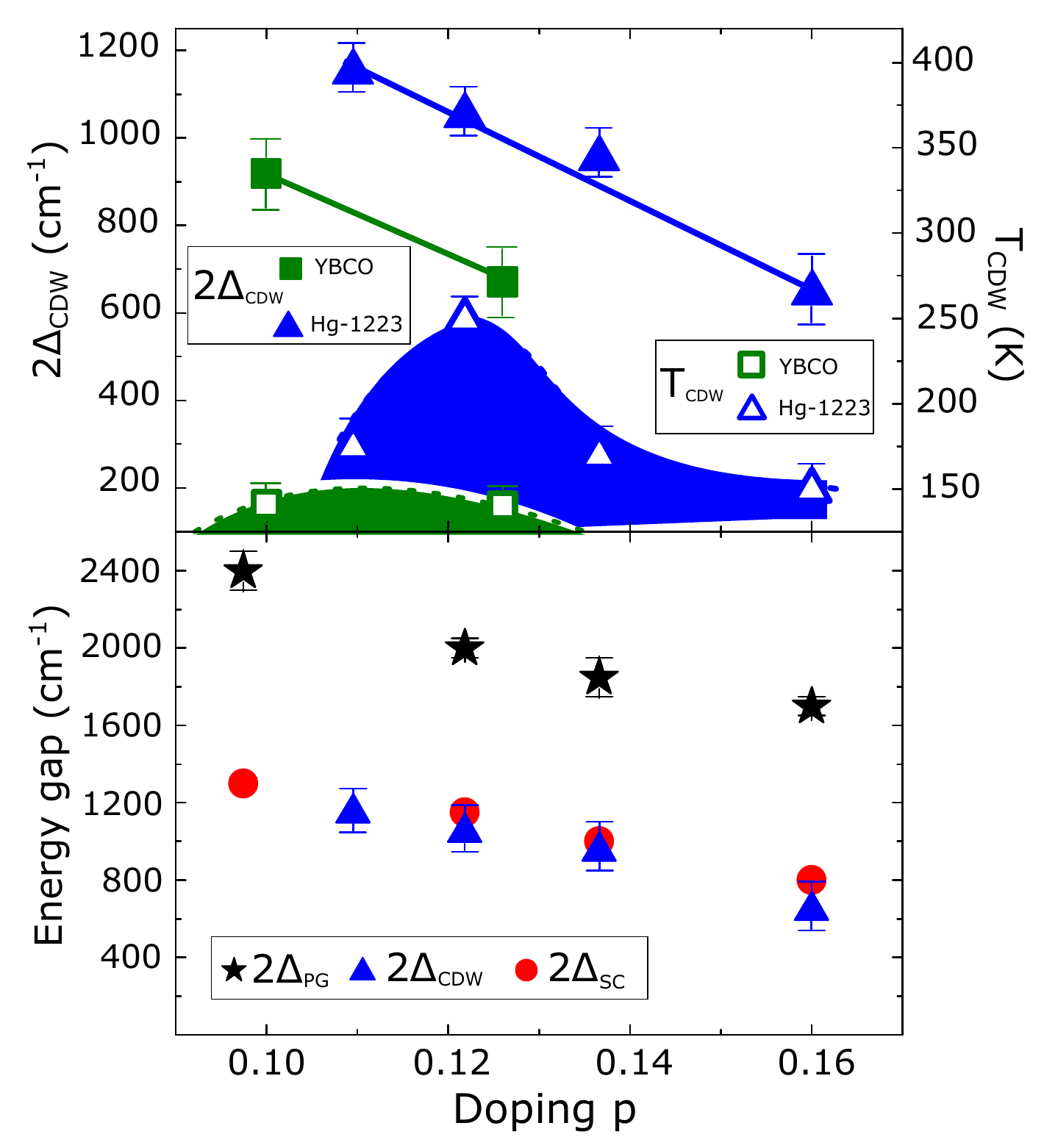}\label{Fig7a}

}

\subfloat[]{\includegraphics[width=8cm]{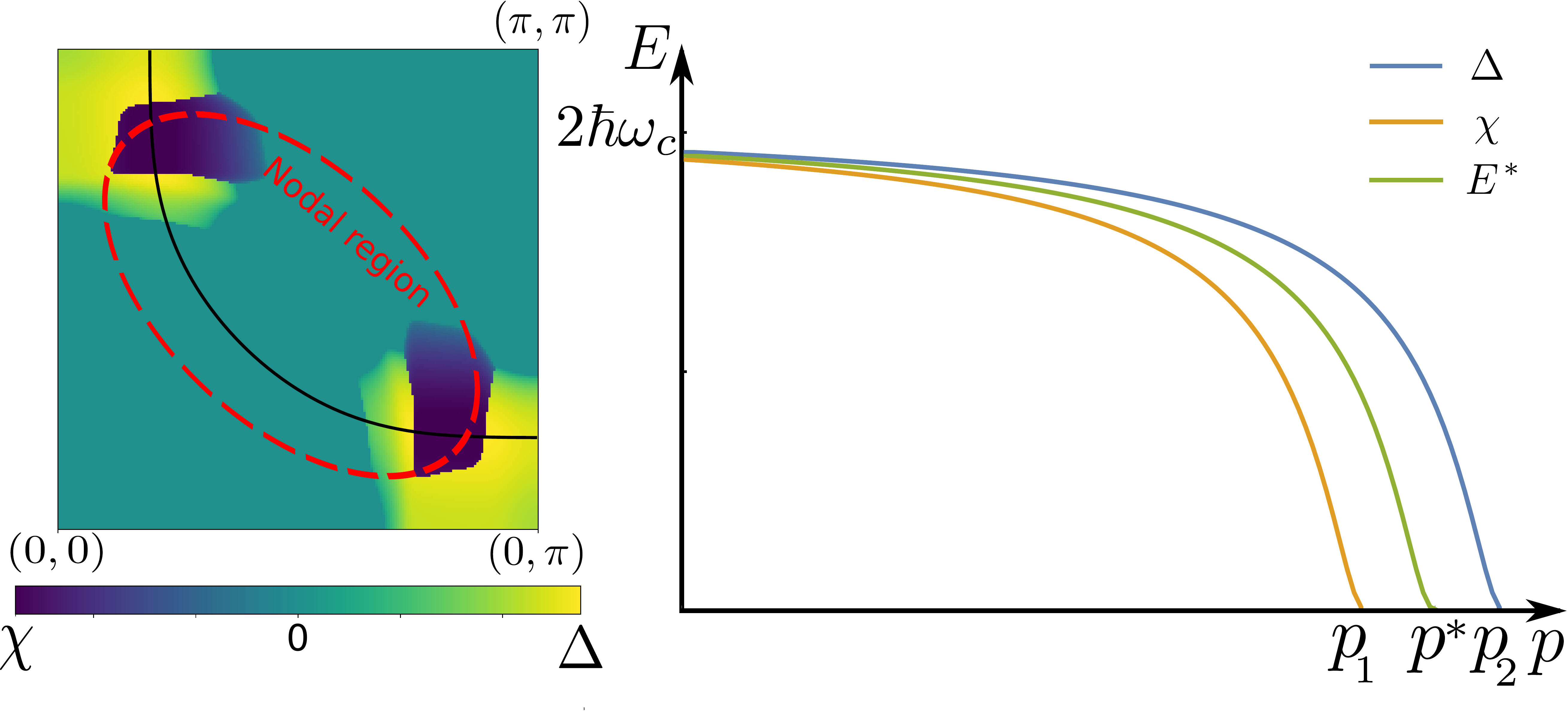}\label{Fig7b}

}

\caption{a) Raman Scattering experiment \cite{Loret19} in Hg1223 where the two spectroscopic
gaps in the $B_{1g}$ (Cooper pairing called $\Delta_{SC}$ here)
and $B_{2g}$ (particle-hole pairing called $\Delta_{CDW}$ here)
channels are shown to be of the same order or magnitude and to decrease
linearly with doping like $T^*$. The higher energy hump $\Delta_{PG}$ also has the same doping dependence and is thus possibly not an independent energy scale. b) Theoretical solution of the gap equations Eq.(\ref{eq:15}) from the toy model Eq.(\ref{eq:8}). In the left
hand side, a quarter of the Brillouin zone is represented with the
Cooper pairing gap $\Delta$ in yellow and the preformed particle-hole
gap $\chi$ in blue. Calculations are done for $T\ge T_{c}$. On the right hand side, the variation of the
magnitude of the gap with doping is given, using a simplified BCS-like scaling argument.}
\end{figure}

A recent Raman scattering experiment performed on the compound Hg1223
shows for the first time that there is a precursor gap in the charge
channel (see Fig.\ref{Fig7a}) \cite{Loret19}. It is seen as a spectral
peak in the $B_{2g}$ channel, which is visible below the ordering
temperature $T_{co}$ but follows $T^{*}$ rather than $T_{co}$ as
a function of oxygen doping. This peak is attributed to charge modulations.
The situation is very similar at what is observed in the AN region
of the Brillouin zone. In the $B_{1g}$ channel (which is scanning
the AN region) a pair breaking peak is observed below $T_{c}$ , but
which follows $T^{*}$ rather than $T_{c}$ with doping. This spectral
peak is also seen in ARPES \cite{Damascelli03,Vishik18}, where it
is shown that it remains unchanged above $T_{c}$ and just broadens
through an additional source of damping \cite{Norman:1998va,Banerjee:2011cu,Vishik18},
up to the PG temperature scale $T^{*}$. It is also seen in STM \cite{Kugler01},
at similar energy scales. A new feature of this
very spectacular Raman scattering experiment is that it answers for
the first time to the question of whether charge modulations are ``strong''
or ``weak'' phenomenon in under-doped cuprates. Charge modulations
have been seen in most of the compounds in the under-doped region,
and although they are bi-dimensional, in theory they are ``long range
enough'' to lead to a re-configuration of the Fermi surface observed
through Quantum Oscillations (QO) \cite{Tabis14}. But in this specific
Raman experiment we see for the first time that the magnitude of the
charge precursor gap in the $B_{2g}$ channel is comparable to the
magnitude of the Pair breaking gap in the $B_{1g}$ channel, comforting
the idea that the charge modulation is a ``strong'' effect in the
physics of underdoped cuprates.\\

At this point, it is important to note that historically in the physics of cuprates, it is argued that there exists two energy scales corresponding to the PG phase (see e.g. the review \cite{Lee06}). First there is a higher energy scale associated with the depletion in the NMR Knight shift, referred to as the ``Alloul-Warren'' gap. This scale is also seen in Raman spectroscopies as a hump rather than a peak. Second scale corresponds to the spectroscopy peaks at a lower energy scale seen in STM, ARPES or Raman spectroscopy. Both the ``Alloul-Warren'' and the spectroscopic gaps seem to be related to each other and behave similarly with doping, decreasing in a quasi-linear fashion as doping increases\cite{Loret19}. The argument of two PG energy scales is typical of strong coupling theories where the higher energy scale is characterized with spin singlet formation and the lower energy scale is associated with superconducting fluctuations. In this work, the PG state is accompanied by a single energy scale $E^*$ which is observed as spectroscopy peaks. The higher energy hump in Raman spectroscopy or the ``Alloul-Warren'' gap is not an independent energy scale and might be related to the coupling of fermions to a collective mode \cite{Eschrig:2000bf,Eschrig:2006ky,Chubukov06}.

Using a BCS-like scaling argument for the different energy
scales, 
\begin{align}
\chi= 2 \hbar \omega_c e^{-{1}/{J_{1}\rho_{0}}},\nonumber \\
\Delta= 2 \hbar \omega_c e^{-{1}/{J_{2}\rho_{0}}},\nonumber \\
E^{*}= 2 \hbar \omega_c e^{-{1}/{J^{*}\rho_{0}}}
\end{align}
where the density of state at the Fermi level $\rho_{0}$ is taken to be constant, $\omega_c$ is the cut-off frequency of the interaction and
$J^{*}=(J_{1}+J_{2})/2$, we can obtain their approximate doping dependence. The evolution of the energy scales
with doping is shown on the right panel of Fig.\ref{Fig7b} where $J_{1}$
and $J_{2}$ have been chosen to be linearly decreasing with doping and vanish at two different doping
$p_{1}$ and $p_{2}$. In this simplified approach, $E^{*}$ goes
to zero at an intermediate doping $p^{*}$. We obtain the linear dependence of $\Delta$ and $\chi$ in an extended range of doping with $\Delta \approx\chi$ as observed in the Raman experiment Fig.\ref{Fig7a}. We also show that there exists only one PG energy scale $E^{*} \approx \Delta \approx\chi$. The
concept of preformed pairs can then account, without resorting to
the idea of a topological order, for the observation that upon application
of a strong magnetic field, the PG survives without a true $T=0$
long range order \cite{Badoux16}. Indeed around $p^{*}$
application of a magnetic field up to $100$T will certainly destroy
SC phase coherence, but to destroy the pairs a much bigger field will
be called for.

\textcolor{black}{Our theory \cite{Chakraborty19} addresses directly
the formation of the spectral gaps $\chi$ and $\Delta$ which are
visible in the $B_{2g}$ and $B_{1g}$ channels respectively 
\begin{align}
\Delta_{k,\omega} & =-\frac{1}{\beta}\sum_{q,\Omega}\frac{J_{-}\left(q,\Omega\right)\Delta_{k+q}}{\left(\omega+\Omega\right)^{2}-\xi_{k+q}^{2}-\Delta_{k+q}^{2}},\label{eq:15}\\
\chi_{k,\omega} & =-\frac{1}{\beta}\sum_{q,\Omega}\frac{J_{+}\left(q,\Omega\right)\chi_{k+q}}{\left(\omega+\Omega-\xi_{k+q}\right)\left(\omega+\Omega-\xi_{k+Q+q}\right)-\chi_{k+q}^{2}}.\nonumber 
\end{align}
with $J_{\pm}\left(q,\Omega\right)$ being related to the original
model parameter as $J_{\pm}\left(q,\Omega\right)\sim3J(p)\pm V$ and
$\beta$ is the inverse temperature.} Solving Eq.(\ref{eq:15})
while keeping the momentum dependence of the gaps and allowing only
one of them to be non-zero at each k-point we obtain the repartition
shown in the left panel of Fig.\ref{Fig7b}. We see that the particle-particle pairing
gap (yellow) is favored in the ANR while the particle-hole pairing gap (violet) prevails in the nodal region. This result agrees well
with the Raman experiment where the SC pair-breaking peak appears in
the $B_{1g}$ symmetry which is probing the anti-nodal part of the Brillouin
zone, while the precursor in the charge channel appears in the $B_{2g}$
symmetry probing the nodal region. The PG energy scale is identified in the present theory as the scale at which the two preformed pairs start to get entangled and compete, $E^{*}=\sqrt{\left|\chi\right|^{2}+\left|\Delta\right|^{2}}$, as first described in Eq.(\ref{eq:10}) for the case of the emergent SU(2) symmetry. Theories which can account for a precursor gap in the charge
channel, which does not follow $T_{co}$ but $T^{*}$ with doping,
and which is of the same size as the precursor gap in the Cooper channel,
are very rare.\textcolor{violet}{{} }\textcolor{black}{This set of experiments,
if confirmed, thus can be considered as a signature of a mechanism
for two kinds of entangled preformed pairs. It has to be noted that
the precursor in the charge channel has been seen, so far, only in
the Hg1223 compound, where the three layers enhance the visibility
of the charge channel. Similar findings were reported for YBCO but
the analysis is more difficult due to the persistent noise in the
$B_{2g}$ channel.}

\subsubsection{Phase locking at $T^{*}$}

The locking of the global phase of the two kinds of preformed pairs
at $T^{*}$ has serious experimental consequences. The PP and PH pairs
become entangled at $T^{*}$ and at lower temperatures $T_{c}$, the remaining relative phase is fixed. Now, if one applies an external
magnetic field, it will create vortices where the SC order parameter
will be suppressed. Inside each vortex the competing order parameter
(in this case a bond excitonic order) is present, which typically
occurs in theories with competing orders. A PDW order will be present
in the vortex halo \cite{Hamidian16,Edkins18,Wang18,Dai18}. But a
stranger feature can be inferred by our theory. Since the phases of
the Cooper pairs and the charge modulations have been locked at $T^{*}$
, inside the vortices, the ``phase-slip'' of the charge modulations
$\cos\left(\mathbf{Q}\cdot\mathbf{r}+\theta_{r}\right)$ is locked
over a very long range $\left\langle \theta_{r}\right\rangle \neq0$,
longer than the typical size of each modulation patch, and thus much
longer than the size of each vortex. This situation was observed in
STM, where one sees that although there is a bit of spreading of the
phase-slips around the mean value, the mean value itself is typically
long range over the whole sample \cite{Hamidian15}. A concrete and
strong experimental prediction would then naturally be to check the
link between the patches of charge modulations and the SC phase. A
Josephson-type of setup, where the SC phase is monitored through
an applied current, would inevitably produce a correlation in the
``phase-slips" of the charge modulations, a situation so unusual that,
if verified, it would probably pin this theory to be the correct one.

\subsubsection{$Q=0$ orders at $T^{*}$}

One of the great experimental complexity of the PG line, is that $\mathbf{Q}=0$
orders have been observed around this line which break discrete symmetries.
Loop currents observed through elastic neutron scattering \cite{Fauque06},
nematicity revealed through magneto-torque measurement \cite{Sato:2017hg,Murayama18}
and parity breaking observed via a second harmonic generation \cite{Zhao2016},
all show a thermodynamic signature at $T^{*}$, the caveat being that
$\mathbf{Q}=0$ orders cannot open a gap in the electronic density
of states. We are thus in a complex situation where the depletion
of the density of states in the ANR cannot be explained by the $\mathbf{Q}=0$
orders whereas any theory which gives an understanding of the PG will
have to account for the presence of these intra unit cell orders at
$T^{*}$.

The phase transition at $T^{*}$ is of a very peculiar nature, with
the freezing of the global phase of the spinor in Eq.(\ref{eq:13})
but also the opening of a gap it looks like a usual Higgs phenomenon.
But the difference with the Meisner effect in a superconductor for
example, is that the opening of the gap is made of a composite order
rather than corresponding to the condensation of a simple field. The
entanglement of two kinds of preformed pairs induces multiple orders
in the PG phase. One can consider a composite field $\phi=\chi\Delta$
as a direct product of the two entangled fields. As the field $\phi$
correspond to particle-particle pairs with a finite center of mass
momentum, this field has the same symmetries as that of a PDW field.
Since $\phi$ has only a global phase and that it is frozen at $T^{*}$
a global phase coherence sets up. But still, the field $\phi$ has
huge fluctuations at $T^{*}$ around a mean average $\left\langle \phi\right\rangle =\left\langle \left|\chi\Delta\right|e^{i\mathbf{Q}\cdot\mathbf{r}}\right\rangle $
in which the fluctuations of the amplitude might ``wash out'' the
modulation term. The PDW field will condense to a long range order
for temperatures below $T_{fluc}$ where both the SC and the bond-excitonic
orders obtain uniform components. One feature of this PDW is that
its modulation wave vector will be same as that of the bond-excitonic
order. Since we work with complex fields $\chi$ and $\Delta$, the
theory can potentially open auxiliary orders around $T^{*}$ especially
those, like loop currents or nematic order, which occur at $Q=0$
but break a discrete symmetry $Z_{2}$ or $C_{4}$. This situation
has been described in previous works \cite{Agterberg:2014wf,Wang15a},
in particular in the case where $T^{*}$ is identified with the formation
of a long range PDW order \cite{Agterberg:2014wf} and where the second
order magneto-electric tensor $l=\left|\phi_{\mathbf{Q}}\right|^{2}-\left|\phi_{-\mathbf{Q}}\right|^{2}$
acquires a non zero value \cite{Sarkar19}.

\section{Conclusion}

We have presented in this paper a review of our understanding of the
PG phase of cuprate superconductors. This is an old problem, but which
has generated a considerable amount of creativity in the last thirty
years, both with theoretical concepts and with new experiments. We
have put forward a scenario for the PG where a phase transition occurs
at $T^{*}$ where two kinds of preformed pairs, in the particle-hole
and particle-particle channels, get entangled and start to compete.
The effective model below $T^{*}$ still retains a large amount of
fluctuations, as is described as an O(3) NL$\sigma$M, or equivalently
the SU(2) chiral model, which is reminiscent of theories of emergent
symmetries with non abelian groups like the SU(2) group. We stress that although the fluctuations
below $T^{*}$ belong to the same universality class as emergent symmetries,
the new mechanism does not rely on the presence of an exact symmetry
in the Lagrangian over the whole under-doped region. The gap opens
in the ANR of the Brillouin zone due to the freezing of the global
phase of our two kinds of preformed pairs. The model has a $U(1)\times U(1)$
gauge structure which enables to identify two phase transitions, one
at $T^{*}$ and another one at lower temperatures, at $T_{c}$. It
has to be noted that models with SU(2) gauge structure, both in the
spin sector and the pseudo-spin sectors have been put forward recently
to explain the PG though corresponding Higgs transitions at $T^{*}$.

The main difference between our approach and those models is that
we do not ``fractionalize'' the electron at $T^{*}$ into ``spinon''
and ``holons'' of some kinds. Instead, the electron keeps its full
integrity, and the PG is due in our scenario to the entanglement of
two kinds of preformed pairs in the charge and Cooper channels. The
two approaches being antinomic, the future will tell whether one of
the two is the right one (or whether maybe a third line of idea is
needed). An argument in our favor lie in the recent observation of
a precursor gap in the charge channel whose energy scale is related
to the PG energy scale \cite{Loret19}. In consideration of the fractionalized
scenario, it has been argued that there is some continuity between
the PG at half filling and the PG in the under-doped region where the
ground state is a superconductor, which would push the balance towards
a fractionalized scenario for $T^{*}$ (see e.g. \cite{Sakai:2018jq}
and references therein). We argue that ``something'' must happen
at a doping of the order of $p\sim0.05$ where numerous of ``bond
orders'' sitting on the Oxygen atoms start to show up. Our intuition
is that it is maybe the doping at which we lose the Zhang-Rice singlet
\cite{Zhang:1988tq}, which thus liberates some degrees of freedom
on the Oxygen atoms, leading to the formation of d-wave ``bond''
charge orders. At this stage it is just a speculation.

The scenario of entangled preformed pairs being simple enough, we hope that it will be possible to produce a strong predictive experiment in the near future.

\textcolor{black}{This work has received financial support from the
ERC, under grant agreement AdG-694651-CHAMPAGNE. The authors also
like to thank the IIP (}Natal, Brazil), for very inspirational visits,
and for wonderful hospitality.

 \bibliographystyle{phaip}
\bibliography{Cuprates}

\end{document}